\documentclass[aps,pra,twocolumn,showpacs]{revtex4}

\usepackage{amssymb}

\begin{document}

\title{The effect of linear terms in a quadratic Hamiltonian}

\author{Mark Andrews}
\email[]{Mark.Andrews@anu.edu.au}

\affiliation{Department of Physics, Australian National University, ACT 0200, Australia}

\date{\today}

\begin{abstract}
For a non-relativistic particle subject to a Hamiltonian that is quadratic in position and momentum, with coefficients that may vary with time, it is shown that the effect of the linear terms in the Hamiltonian is just a spatial translation of the wave function and a change in its phase. The shifts in position and phase can be expressed in terms of classical trajectories. This simple effect of the linear terms is related to the fact that all moments about the centroid of the wave function evolve independently of the linear terms.
\end{abstract}


\maketitle

\section{Introduction}

Quadratic Hamiltonians have special properties and the evolution of wave functions is closely related to classical mechanics. For example, any time dependence in the Hamiltonian can be removed by a linear canonical transformation and the propagator can be expressed in terms of solutions of ordinary differential equations related to those of the classical motion.\cite{QHams} Ehrenfest's theorem shows that the centroid $\langle \hat{x}\rangle$ of any wave packet will exactly follow a classical trajectory. 

The subject of the present paper is the effect of the linear terms in quadratic Hamiltonians, \textit{i.e.} terms linear in position or momentum. A term linear in position represents a spatially uniform force, such as gravity or a uniform electric field. Bowman\cite{Bow} has shown that the effect of such a term on the wave function of an otherwise free particle is just a spatial displacement and a shift in the phase. Here we extend that work to show that there is just a shift in position and phase due to the linear terms in any quadratic Hamiltonian, with arbitrary time-dependence in the coefficients in the Hamiltonian. Thus, for example, if an harmonic oscillator is subject to a force that does not vary with position but may vary with time, then the wave function can be simply derived from the wave function without the extra force.

That there should be a simple relation between the wave functions with and without the linear terms was strongly suggested by the result, discussed in Section IV, that all moments relative to the centroid evolve independently of any linear terms in the Hamiltonian. It is hard to believe that this could happen without some simple relation between the wave functions.

Ehrenfest's theorem prescribes how $\langle \hat{x}\rangle$ and $\langle \hat{p}\rangle$ are affected by the linear terms and we use this to determine the shift in position and the shift in phase apart from an added term $\beta(t)$, which is then determined by inserting the trial form of the wave function into Schr\"odinger's equation.

\section{Transforming Schr\"odinger's equation}

The quadratic Hamiltonian has the form
\begin{equation}
\label{eq:Ham}
\hat{H}(\hat{p},\hat{x})=\frac{1}{2}a\hat{p}^{2}+\frac{1}{2}b(\hat{p}\hat{x}+\hat{x}\hat{p})+\frac{1}{2}c\hat{x}^{2}+f\hat{p}+g\hat{x},
\end{equation} 
where the coefficients $a, b, c, f, g$ are real and may depend on the time. The classical equations of motion are
\begin{equation}
\label{eq:motionClass}
d_{t}x=ap+bx+f,\,\,\,\,\,-d_{t}p=bp+cx+g.
\end{equation}
The spatial translation due to the linear terms, $f$ and $g$, is the same as the change in position of a classical particle. Due to the linearity of the equations of motion, the differences $\bar{x}$ and $\bar{p}$ between a classical trajectory with $f$ and $g$ and one without will also satisfy Eq.(\ref{eq:motionClass}). There can be no difference between the trajectories at time $t_{0}$ when $f$ or $g$ are turned on. Therefore the shifts $\bar{x}$ and $\bar{p}$ required are the solutions to Eq.(\ref{eq:motionClass}) with $\bar{x}(t_{0})=0$ and $\bar{p}(t_{0})=0$. The change $\bar{p}$ in momentum will be related to the change in phase. If we assume that the wave function has the form $\psi(x,t)=\exp[\imath\,\theta(x,t)]\,\Psi(x-\bar{x},t)$, then $\langle \psi|\hat{p}|\psi\rangle=\hbar\,\partial_{x}\theta+\langle\Psi|\hat{p}|\Psi\rangle$ and therefore $\hbar\theta=\bar{p}x-\beta(t)$. Hence we insert a wave function of the form
\begin{equation}
\label{eq:Psi}
\psi(x,t)=\exp[\frac{\imath}{\hbar} \big(\bar{p}(t)x-\beta(t)\big)]\,\Psi(\xi,t),
\end{equation}
where $\xi=x-\bar{x}(t)$, into Schr\"odinger's equation
\begin{equation}
\label{eq:Sch}
(-\imath\hbar\partial_{t}+\hat{H})\psi=0\,\,\,\,\,\text{with}\,\,\,\,\,\hat{p}=-\imath\hbar\partial_{x}.
\end{equation}
The result is that
\begin{eqnarray}\label{eq:Sch2}
[-\imath\hbar\partial_{t}+\hat{H}(\hat{p}_{x},x)]e^{\imath(\bar{p}x-\beta)/\hbar}\Psi(x-\bar{x},t)= \\ \nonumber
e^{\imath(\bar{p}x-\beta)/\hbar}[-\imath\hbar\partial_{t}+\hat{H_{0}}(\hat{p}_{\xi},\xi)]\Psi(\xi,t),
\end{eqnarray}
where $\hat{H}_{0}$ is the Hamiltonian without the linear terms, provided that $d_{t}\beta=\frac{1}{2}a\bar{p}^{2}-\frac{1}{2}c\bar{x}^{2}+f\bar{p}$. However, from Eq.(\ref{eq:motionClass}), $d_{t}(\bar{p}\bar{x})=a\bar{p}^{2}-c\bar{x}^{2}+f\bar{p}-g\bar{x}$ and therefore
\begin{equation}
\label{eq:beta}
\beta(t)=\frac{1}{2}\bar{p}(t)\bar{x}(t)+\frac{1}{2}\int^{t}_{t_{0}}[f(t')\bar{p}(t')+g(t')\bar{x}(t')]dt'.
\end{equation}
Thus, we have shown that 
\begin{equation}
\label{eq:psi}
\psi(x,t)=\exp[\frac{\imath}{\hbar} \big(\bar{p}(t)x-\beta(t)\big)]\,\Psi(x-\bar{x},t)
\end{equation}
will satisfy Schr\"odinger's equation with Hamiltonian $\hat{H}$ if $\Psi(x,t)$ satisfies Schr\"odinger's equation with Hamiltonian $\hat{H}_{0}$. Since $\Psi(x,t_{0})=\psi(x,t_{0})$, the effect of the linear terms in the Hamiltonian is just a shift of $\bar{x}$ in position and a shift of $(\bar{p}x-\beta)/\hbar$ in phase.

\vspace{2mm}
\textbf{Example 1.} \textit{A particle with a uniform force.} If a uniform force $-g(t)$ is turned on at time $t=t_{0}$ then the wave function for $t>t_{0}$ is
\begin{equation}
\label{ }
\psi(x,t)=\exp[-\frac{\imath}{\hbar}(G(t)x+\frac{G_{2}(t)}{2m})\Psi(x+\frac{G_{1}(t)}{m},t)
\end{equation}
where $G(t)=\int_{t_{0}}^{t}g(t')dt'$, $G_{1}(t)=\int_{t_{0}}^{t}G(t')dt'$, $G_{2}(t)=\int_{t_{0}}^{t}G(t')^{2}dt'$ and $\Psi(x,t)$ is the wave function with no force, \textit{i.e.} the wave function for a free particle.

\vspace{2mm}
\textbf{Example 2.} \textit{An oscillator subject to a uniform force.} If a uniform force $-g(t)$ is turned on at time $t=t_{0}$ then the wave function for $t>t_{0}$ is
\begin{equation}
\label{ }
\psi(x,t)=\exp[-\frac{\imath}{\hbar}\big(C(t)x+\beta(t)\big)]\Psi(x-\frac{S(t)}{m\omega},t)
\end{equation}
where $\Psi(x,t)$ is the wave function for the unforced oscillator, $S(t)=\int_{t_{0}}^{t}g(t')\sin \omega(t-t')dt'$, $C(t)=\int_{t_{0}}^{t}g(t')\cos \omega(t-t')dt'$, and
\begin{equation}
\label{ }
\beta(t)=\frac{1}{2m\omega}[S(t)C(t)-\int_{0}^{t}g(t')S(t')\,dt'].
\end{equation}

\section{\label{sec:Momentum}Momentum wave function}

In view of the similarity in the roles of position and momentum in the Hamiltonian, it is to be expected that the momentum wave function will also be affected by the linear terms only through a displacement (in momentum) and a phase shift. The momentum wave function corresponding to the wave function $\psi(x,t)$ is
\begin{equation}
\label{ }
\phi(p,t)=\frac{1}{\sqrt{2\pi\hbar}}\int_{-\infty}^{\infty}\exp(-\frac{\imath}{\hbar}px )\psi (x,t)\,dx.
\end{equation}
Inserting $\psi$ from Eq.(\ref{eq:psi}) and rearranging the exponent,
\begin{eqnarray} \nonumber
\phi(p,t) & = & \frac{e^{-\imath(\bar{x}p-\gamma)/\hbar}}{\sqrt{2\pi\hbar}}\int_{-\infty}^{\infty}\exp[-\frac{\imath}{\hbar}(p-\bar{p})\xi]\Psi (\xi,t)\,d\xi \\
 & = & e^{-\imath(\bar{x}p-\gamma)/\hbar}\Phi(p-\bar{p},t),
\end{eqnarray}
where $\Phi(p,t)$ is the momentum wave function without the linear terms and
\begin{equation}
\label{eq:beta}
\gamma(t)=\frac{1}{2}\bar{p}(t)\bar{x}(t)-\frac{1}{2}\int^{t}_{t_{0}}[f(t')\bar{p}(t')+g(t')\bar{x}(t')]dt'.
\end{equation}

\section{\label{sec:Moments}Moments ignore linear terms}

The simplest moments (relative to the centroid) are the second order ones: $\Delta_{x}^{2}=\langle (\hat{x}-\langle \hat{x} \rangle)^{2}\rangle$, $\Delta_{p}^{2}=\langle (\hat{p}-\langle \hat{p} \rangle)^{2}\rangle$ and the correlation $\Delta_{xp}=\langle \hat{p}\hat{x}+\hat{x}\hat{p}\rangle-2\langle \hat{p}\rangle\langle\hat{x} \rangle$. But there are an infinite number of higher moments involving expectation values of higher powers of $\hat{x}-\langle \hat{x} \rangle$ and $\hat{p}-\langle \hat{p} \rangle$ and products of these. That  $\langle (\hat{x}-\langle \hat{x} \rangle)^{n}\rangle$ is independent of $f$ and $g$ follows easily from the form of the wave function in Eq.(\ref{eq:psi}): $\langle\psi|(x-\langle{x}\rangle)^{n}|\psi\rangle=\langle\Psi|(\xi-\langle{\xi}\rangle)^{n}|\Psi\rangle$. Similarly for $\langle (\hat{p}-\langle \hat{p} \rangle)^{n}\rangle$, using the form of the momentum wave function. The moments that involve both $\hat{x}$ and $\hat{p}$ are not so simple, and here the term $\bar{p}x/\hbar$ in the phase is important. First verifying that $(\hat{p}_{x}-\langle \hat{p} \rangle_{\psi})^{n}\psi=\exp[\imath(\bar{p}x-\beta)/\hbar](\hat{p}_{\xi}-\langle \hat{p} \rangle_{\Psi})^{n}\Psi$ makes the calculation straightforward.

The evolution of these moments can be analyzed in a completely different way. For the quadratic Hamiltonian in Eq.(\ref{eq:Ham}), the Heisenberg equations of motion have the same form as the corresponding classical equations:
\begin{equation}
\label{eq:opMotion}
d_{t}\hat{x}=a\hat{p}+b\hat{x}+f,\,\,\,\,\,-d_{t}\hat{p}=b\hat{p}+c\hat{x}+g,
\end{equation}
[In Schr\"odinger's picture, the total time derivative\cite{A} of any operator $\hat{A}$ is $d_{t}\hat{A}=\partial_{t}\hat{A}+\imath \hbar^{-1}[\hat{H},\hat{A}]$ and Eq.(\ref{eq:opMotion}) follows. Then $d_{t}\langle \hat{A}\rangle =  \langle d_{t} \hat{A}\rangle$ and therefore the expectation values of position and momentum follow a classical trajectory, which is Ehrenfest's result for this system.]

For the moments, we need the deviations from the expectation values. Thus we introduce the operators $\hat{X}=\hat{x}-\langle \hat{x}\rangle$ and $\hat{P}=\hat{p}-\langle \hat{p}\rangle$ and then
\begin{equation}
\label{eq:inhom}
d_{t}\hat{X}= a\hat{P}+b\hat{X},\hspace{3mm}-d_{t}\hat{P}=b\hat{P}+c\hat{X}.
\end{equation}
Thus the linear terms in the Hamiltonian are absent from the equations of motion for the deviations. This implies that the moments evolve in the same way whether the linear terms are present or not. 

To see how equation (\ref{eq:inhom}) determines the evolution of the moments, consider the second-order moments:
\begin{eqnarray}
d_{t}\hat{X}^{2}& = &a(\hat{P}\hat{X}+\hat{X}\hat{P})+2b\hat{X}^{2} \\
d_{t}(\hat{P}\hat{X})& = &a\hat{P}^{2}-c\hat{X}^{2}\\
d_{t}\hat{P}^{2}& = &-2b\hat{P}^{2}-c(\hat{P}\hat{X}+\hat{X}\hat{P}).
\end{eqnarray}
The expectation values of these equations then gives
\begin{eqnarray}
d_{t}\Delta_{x}^{2}& = &2a\Delta_{xp}+2b\Delta_{x}^{2} \\
d_{t}\Delta_{xp}& = &a\Delta_{p}^{2}-c\Delta_{x}^{2}\\
d_{t}\Delta_{p}^{2}& = &-2b\Delta_{p}^{2}-2c\Delta_{xp}.
\end{eqnarray}
This closed set of equations can be solved\cite{A1} for $\Delta_{x}^{2}$, $\Delta_{xp}$, and $\Delta_{p}^{2}$ in terms of their initial values, using a basis of solutions of the classical equations of motion (without the linear terms in the Hamiltonian), thus confirming that these moments evolve independently of the linear terms. This approach can be extended to the higher moments.\cite{AH}

\section{\label{sec:Conc}Conclusion}

For quadratic Hamiltonians, dealing with any linear terms is a simple matter: their only effect is to change the position (by the same amount as for a classical particle) and to change the phase. The change in phase is linear in position. The momentum wave function also suffers only the classical shift in momentum and a change in phase (linear in the momentum). These results are consistent with the fact that all moments relative to the centroid evolve independently of the linear terms in the Hamiltonian.

It is well known\cite{QHams,A1} that, for any quadratic, time-dependent Hamiltonian, there are families of Gaussian and Hermite-Gaussian wave packets that retain the shape of $|\psi|^{2}$ as they evolve, though they do change scale. In general other wave packets do change shape (and scale), but our result shows that they have the same shape and scale that they would have without the linear terms.

The discussion given here for one spatial dimension can easily be extended to higher dimensions provided the second-order terms in the Hamiltonian are separable in Cartesian coordinates.

\end{document}